\documentclass[a4paper]{article}
\usepackage{18lomcon}         
\usepackage{cite}             
\usepackage{epsfig}           
\usepackage{epstopdf}
\usepackage{amssymb}
\usepackage{verbatim}

\usepackage{color}
\definecolor{slateblue}{rgb}{0.2,0.2,0.6}
\usepackage[breaklinks=true,colorlinks=true,linkcolor=slateblue,citecolor=slateblue,urlcolor=slateblue,pdfauthor={Tomassetti},pdftitle={TimeLag}]{hyperref}

\usepackage{xspace}
\newcommand{\GCR}{CR\xspace}
\newcommand{\GCRs}{CRs\xspace}

\newcommand{\AMS}{AMS}

\newcommand{\pHe}{\textrm{p/He}}
\newcommand{\p}{\ensuremath{p}}
\newcommand{\He}{\textrm{He}}
\newcommand{\epratio}{{e\ensuremath{^{+}}}/{e\ensuremath{^{-}}}}
\newcommand{\pbarp}{\textsf{\ensuremath{\bar{p}/p}}}
\newcommand{\pbar}{\textsf{\ensuremath{\bar{p}}}}
\newcommand{\eplus}{\textsf{\ensuremath{e^{+}}}}
\newcommand{\eminus}{\textsf{\ensuremath{e^{-}}}}

\newcommand{\etal}{et alii}
 
\newcommand{\ie}{\textit{i.e.}} 

\begin{document}


\title{Solar Modulation of Galactic Cosmic Rays:\\Physics Challenges for AMS-02}
\author{Nicola Tomassetti \email{nicola.tomassetti@cern.ch}}
\affiliation{Universit{\`a} degli Studi di Perugia \& INFN-Perugia, I-06100 Perugia, Italy\\[0.15cm]}
\date{}
\maketitle

%
\begin{abstract} 
  The Alpha Magnetic Spectrometer (AMS) is a new generation high-energy physics experiment
  installed on the International Space Station in May 2011 and operating continuously since then. 
  Using an unprecedently large collection of particles and antiparticles detected in space,
  AMS is performing precision measurements of cosmic ray energy spectra and composition.
  In this paper, we discuss the physics of solar modulation in Galactic cosmic rays that can be investigated
  with AMS my means of dedicated measurements on the time-dependence of cosmic-ray proton, helium, electron and positron fluxes. 
\end{abstract}

\section{Introduction} 
%
The Alpha Magnetic Spectrometer (AMS) is a state-of-the-art particle physics
experiment operating on the International Space Station (ISS) since May 2011.
In the first 6 years of missions, AMS has detected over 100 billion cosmic ray (\GCR) particles.
Recently, it has released new results on proton, antiproton, lepton, and nuclei energy spectra
at unexplored energies and with an unmatched level of
accuracy \cite{Aguilar2016PbarP,Aguilar2016BC,Aguilar2015Proton,Aguilar2015Helium,Aguilar2013PositronFraction}.
A collection of these data is available on the ASI/SSDC-CR database \cite{DiFelice2017}.
The long duration of mission, planned to last for the whole ISS lifetime, will cover a complete solar cycle
from the ascending phase of cycle 24, through its maximum, and the descending phase into the next solar minimum. 
This makes AMS an excellent multichannel \GCR monitor of solar activity \cite{Bindi2017,DellaTorre2016}. 
Precision measurements of the \GCR{} time evolution, in connection with the changing solar activity,
may give us strong insight on the so-called \textit{solar modulation} effect. 

Solar modulation is a time-, space-, energy-, and particle-dependent phenomenon that arises
from basic transport processes of \GCRs in the heliosphere \cite{Potgieter2013}.
Along with its connection with solar and CR physics, understanding CR modulation addresses a prerequisite
for modeling space weather effects, which is an increasing concern for space missions and air travelers.
The study of these effects has been limited for long time by the scarcity of long-term \GCR data on different species, and by the poor knowledge of the local interstellar spectra (LIS).
The very first LIS measurements on the \GCR fluxes have been recently provided by Voyager-1 probe in the interstellar space.
A continuous stream of time-resolved and multichannel CR data is being provided by the AMS on the ISS.
In light of these milestones, new objectives of low-energy physics investigation can be devised:
(i) to advance solar modulation observations of \GCR particles and antiparticles, and
(ii) to develop improved and measurement-validated models of \GCR transport in the heliosphere.

\section{Physics challenges in cosmic-ray modulation} 
%
We now discuss some important physics topics that can be investigated with dedicated analysis of the AMS data on \GCR{} modulation.
The first one deals with puzzling anomalies detected in the energy spectra of CR proton and helium nuclei. 
These spectra are found to harden at rigidity $R=pc/Ze \gtrsim$\,200 GV,
while their \pHe{} ratio as function of rigidity falls off steadily as \pHe\,$\propto R^{-0.08}$ \cite{Aguilar2015Helium}.
The decrease of the \pHe{} ratio is usually interpreted in terms of particle-dependent acceleration, 
but this interpretation is in contrast with the \emph{universal} nature of diffusive-shock-acceleration mechanisms,
\ie, the conception that CRs are injected in the Galaxy by composition-blind rigidity-dependent accelerators.
A model based on two-source components has been proposed in \cite{Tomassetti2015PHeRatio} and
further investigated in \cite{Zhang2017}. In this model, the \pHe{} anomaly is explained by a flux transition
between two source components ($L$-type and $G$-type) having by different injection spectra and composition.
The universality of particle acceleration is not violated in this model,
since each class of source is assumed to provide elemental-independent acceleration spectra. 
A possible signature of this scenario is a progressive flattening of the \pHe{} ratio
at multi-TeV energies, \ie, where the  $G$-type source dominates the \GCR flux.
Such a signature is hinted at by the recent data, but the situation is unclear.
Interestingly, a complementary observational test can be done in the sub-GeV energy regime,
\begin{figure*}[!th]
\begin{center}
\includegraphics[width=0.47\columnwidth]{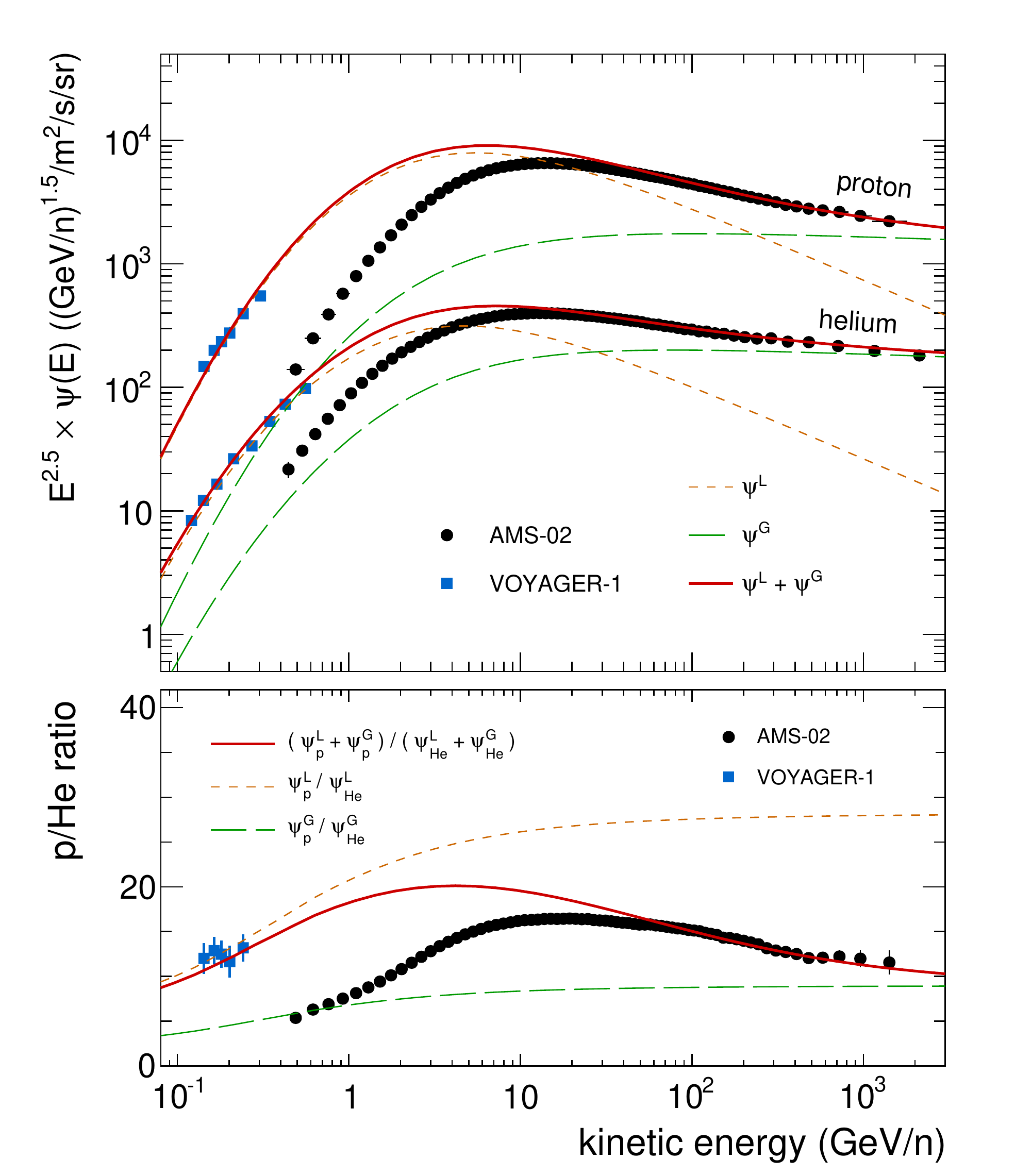}
\includegraphics[width=0.47\columnwidth]{ccLomCon18Fig1a.pdf}
\end{center}
\caption{
  Proton and helium energy spectra (top) and \pHe{} ratio (bottom) from \cite{Tomassetti2017ASR}.
  The calculations (solid lines) are shown in comparisons with the AMS and Voyager-1 data \cite{Aguilar2015Proton,Aguilar2015Helium,Cummings2016}.
  The dashed lines are the single source components.
  Calculations are shown for LIS (left) and modulated (right) fluxes. Solar modulation uncertainties are shown as shaded bands.
}
\label{Fig::ccProtonHeliumMOD}
\end{figure*}
\ie, where the $L$-type source is expected to dominate the flux. 
However, in this energy window, the \GCR fluxes are significantly affected by the time-dependent
effect of solar modulation, hence a careful modeling of the \GCR transport in the heliosphere is essential.
In Fig.\,\ref{Fig::ccProtonHeliumMOD}, from \cite{Tomassetti2015PHeRatio}, it is shown that the  \p-\He data reported by \AMS and Voyager-1
are in good agreement with the model predictions, once the modulation effect is accounted,
therefore supporting the universality of \GCR acceleration mechanism.
However, this study was limited by simple model approach and by 
the lack of time-dependent measurements on \GCR in the period of investigation. 
These measurements are being carried out by AMS over its first 6 years of mission.
This can provide a conclusive observational test
for such a scenario of \GCR{} origin and propagation.
\begin{figure*}[!t]
\centering
\includegraphics[width=0.92\textwidth]{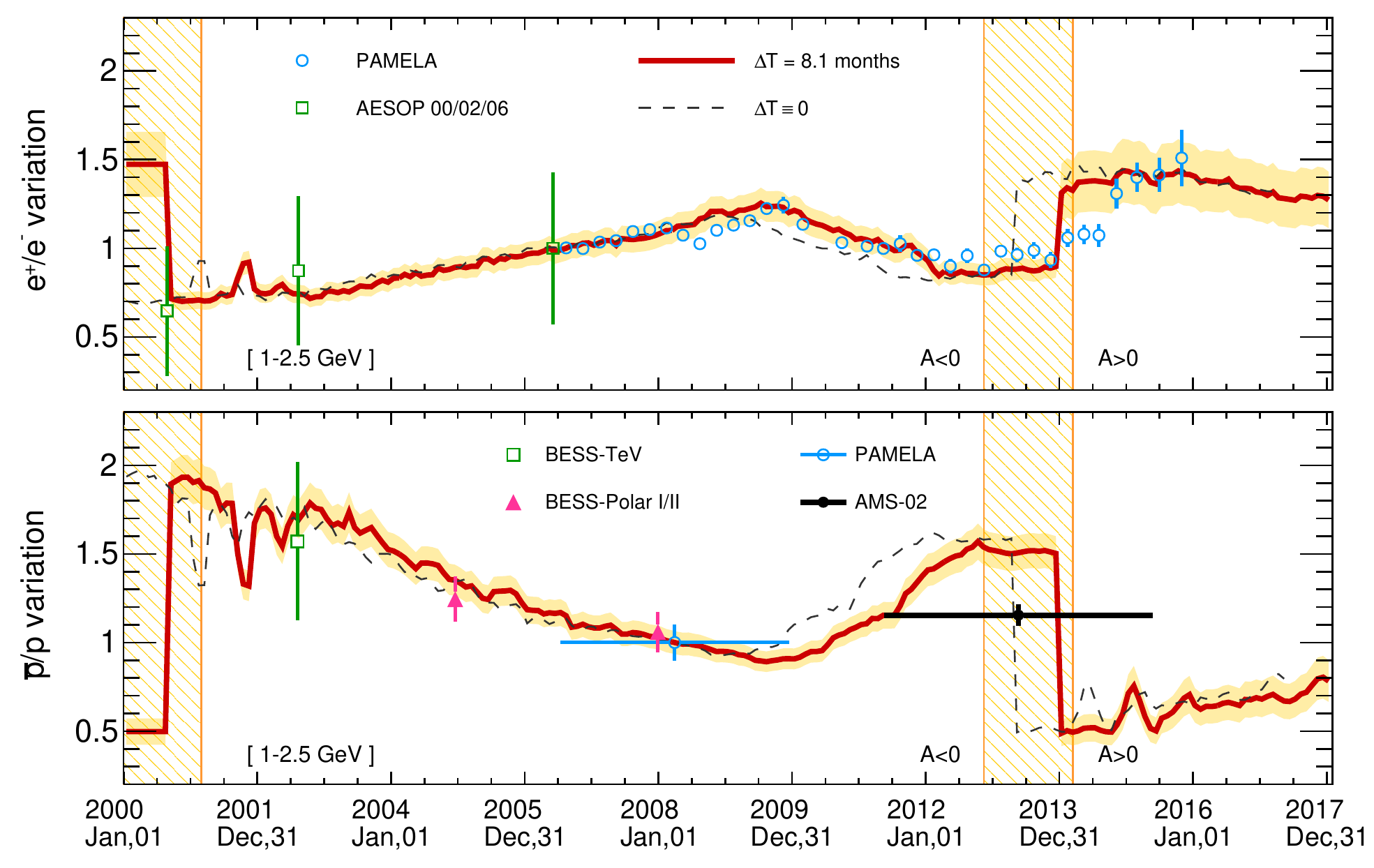}
\caption{\footnotesize{%
Time profile of the ratios \epratio{} (top) and \pbarp{} (bottom) at $E=1\--2.5$\,GeV \cite{Tomassetti2017TimeLag}.
Model predictions and their corresponding uncertainties are shown 
in comparisons with the data \cite{Adriani2016,Aguilar2016PbarP,Abe2012}. 
The shaded bars indicate the magnetic reversals of the Sun's polarity. 
}}
\label{Fig::ccAntimatterMatterRatios}
\end{figure*}

Another important topic is the recent observation of a eight-month \emph{time lag} in solar modulation of \GCR{} \cite{Tomassetti2017TimeLag}.
This effect reveals important properties on the dynamics of the formation and changing conditions of the heliospheric plasma.
While the analysis of \cite{Tomassetti2017TimeLag} is based on \GCR proton,
the key observational tests for the model are given for the evolution
of \GCR{} antimatter/matter ratios, shown in Figure\,\ref{Fig::ccAntimatterMatterRatios}.
Crucial tests can be performed by \AMS{} via monthly-resolved measurements of these ratios,
or even better, by measurements of individual particle fluxes for \p, \pbar, \eplus, and \eminus{}
under both polarity conditions and across the magnetic reversal. 
This demonstrates that time-dependent measurements on CR antimatter can
provide precious information on the physics of the heliosphere.

Understanding charge-sign dependence of CR modulation is also essential to
search for dark matter signatures in CR antimatter fluxes.
Other problems related to uncertainties in solar modulation 
are studied in recent works \cite{Bindi2017,Tomassetti2012Isotopes,Tomassetti2017BCUnc}.
Without dedicated data on the time-dependence of the CR flux, solar modulation models
show degeneracies with the CR propagation parameters and fall short of predicting
the level of astrophysical antimatter background.

\section{Conclusions} 
%
We have briefly discussed some physics challenges, in modern astrophysics, that will benefits from
a dedicated \emph{multichannel investigation of solar modulation effects in Galactic \GCRs}:
the origin of the anomalies in the spectra of CR nuclei,
the dynamics of the changing conditions of the heliospheric plasma,
the parameters constraints in astrophysical models of CR propagations,
the origin of CR antimatter and its connection with the dark matter puzzle.
In order to develop reliable and data-driven models of CR modulation, however, 
the availability of time-resolved measurements over the period of interest is crucial. 
In this respect, monthly-resolved data from AMS will be very precious.

\section*{Acknowledgments} 
%
This project has received funding from the European Union’s Horizon 2020 research and innovation programme under the Marie Sklodowska-Curie grant agreement No 707543 - MAtISSE.


\end{document}